\documentstyle{aipproc-011796}
\begin{document}


\def \ns{\enspace}
\def \ts{\thinspace}
\def \nts{\negthinspace}
\def \beq{\begin{equation}}
\def \eeq{\end{equation}}
\def \beqa{\begin{eqnarray}}
\def \eeqa{\end{eqnarray}}
\def \half{\hbox{$1\over2$}}
 

\def \trans#1{\vec{#1}}
\def \xf{x_F}
\def \LQCD{\Lambda_{\rm QCD}}
\def \Jt{\trans{J}}
\def \bt{\trans{b}}
\def \pt{\trans{p}}
\def \qt{\trans{q}}
\def \xit{\trans\xi}
\def \xiprimet{\xit^\prime}
\def \deltat{\trans\Delta}
\def \sigmat{\trans\Sigma}
\def \delt{\trans\nabla}
\def \xt{\trans{x}}
\def \xprimet{\xt^\prime}
\def \At{\trans{A}}
\def \curlyL{{\cal L}}
\def \curlyD{{\cal D}}
\def \gnum{ {{dN}\over{dq^{+} d^2\qt}} }
\def \cfun{\langle A_i^a(x^{-},\xt) A_j^b(x^{\prime -},\xprimet)\rangle}
\def \TRcfun{\langle A_i^a(x^{-},\xt) A_i^a(x^{\prime -},\xprimet)\rangle}
\def \Ihat{C}
\def \MV{MV}


\title{Gluons in a Color-Neutral Nucleus\footnote{Talk presented
at MRST'99:  High Energy Physics at the Millenium, Carleton University,
Ottawa, Ontario, Canada, May 10-12, 1999.}}

\author{Gregory Mahlon}
\address{Department of Physics, McGill University\\
3600 University St., Montr\'eal, QC, Canada H3A 2T8}

\maketitle

\begin{abstract}
We improve the McLerran-Venugopalan model~\cite{papers1to4,paper9} 
by introducing a charge-density correlation function which is
consistent with the observation that nucleons carry no net
color charge.  The infrared divergence in the transverse
coordinates that was present in the McLerran-Venugoplan 
model is eliminated by the enforcement of color neutrality.
\end{abstract}


The problem of extracting first principles predictions from
the theory of strong interactions,
quantum chromodynamics (QCD), is notoriously difficult, largely
because of the nonlinearity of the theory.  The quanta of the
gauge field, the gluons, themselves carry color charge and so
serve as a source of additional gluons.  Thus, any regime
in  which it is possible to actually
compute some physical observable within the framework of QCD is of 
great interest.

Such is the case with the McLerran-Venugopalan (MV) 
model~\cite{papers1to4,paper9}.  What McLerran
and Venugopalan realized is that for very large nuclei at
very small values of the longitudinal momentum fraction,
the number of color charges participating in the generation
of the QCD vector potential is large.
In this situation, the gluon number density
may be approximated to lowest order by solving the {\it classical}\
Yang-Mills equations in the presence of the (classical)
source generated by the valence quarks.  To actually extract
the gluon number density, we must average over the sources
to obtain the two point correlation function for the vector
potential.  This approximation may be systematically
improved by including the quantum corrections.
The correlation function derived in Ref.~\cite{paper9} is
highly infrared divergent at large transverse distances.
In this talk, we will show that this difficulty may be
ameliorated by forcing the nucleons to obey a color-neutrality
condition\cite{US}.

Before proceeding with the main part of this talk, it is
necessary to say a few words about our notation and conventions.
We elect to work with light-cone coordinates, defined by 
$ x^{\pm} \equiv (x^0 \pm x^3)/\sqrt{2}$.
The components of a 4-vector will be written as
$x = (x^{+},x^{-},\xt)$, with vectors in the two-dimensional
(transverse) subspace written with arrows.
We choose a metric with the signature $(-,+,+,+)$.  Consequently,
we have $x_\pm = -x^{\mp}$, and a dot product which reads
$x \cdot q \equiv - x^{+} q^{-} - x^{-} q^{+} + \xt\cdot\qt$.
We will work in the light-cone gauge, $ A^{+} \equiv 0$,
where the intuitive parton model is realized~\cite{paper31}.

\begin{figure}[t!] 
\includegraphics{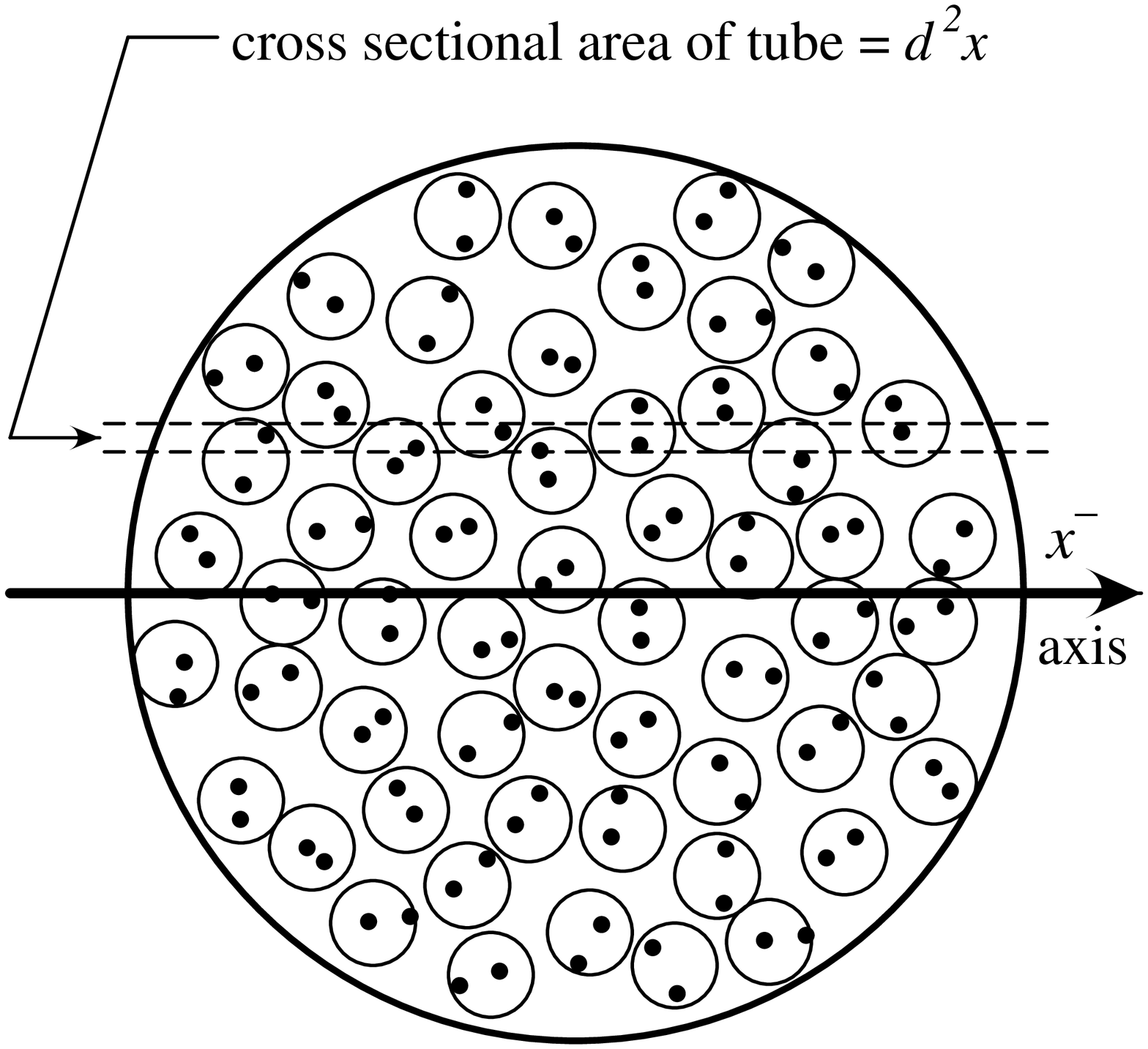}
\includegraphics{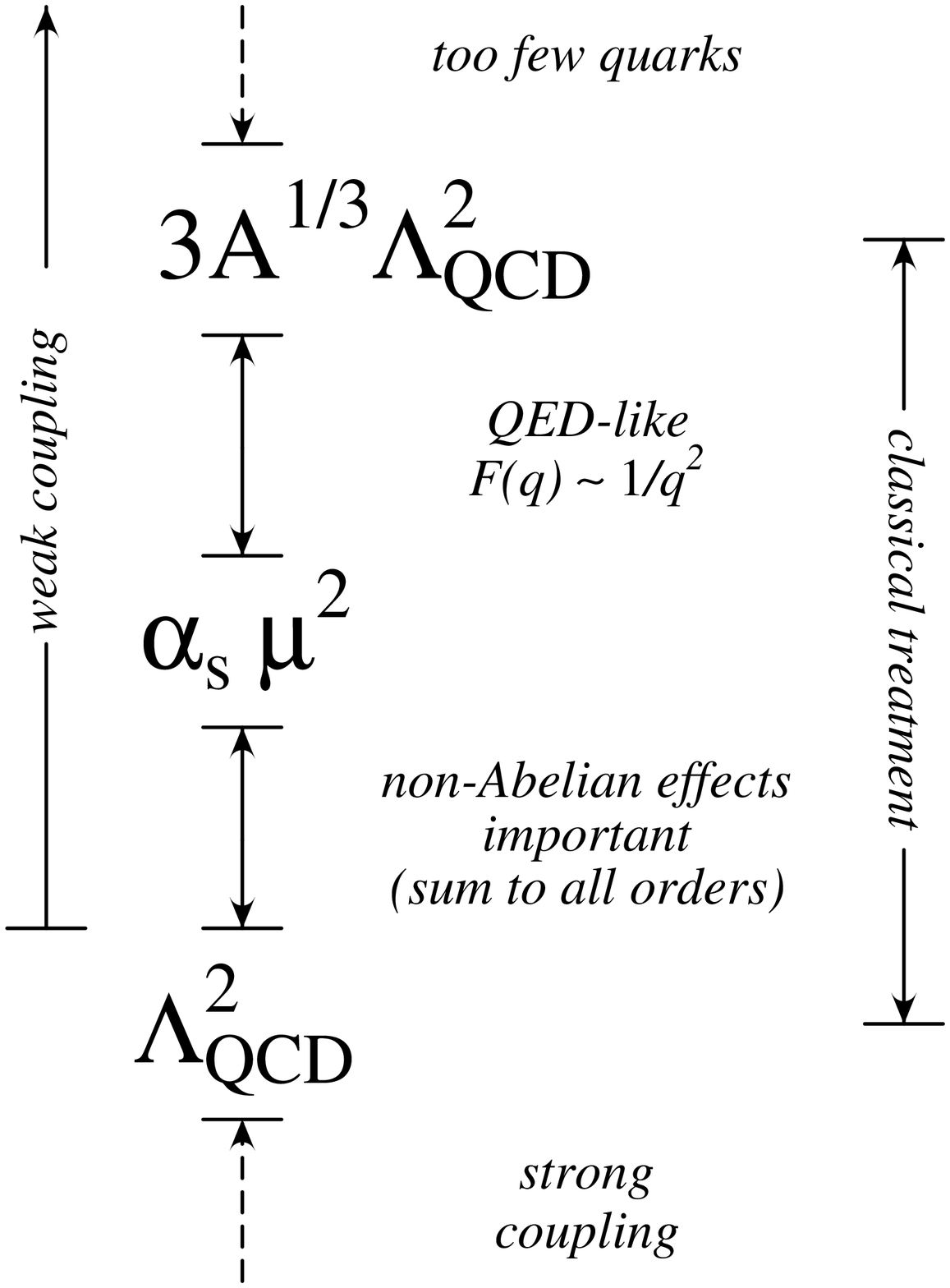}
\vspace{2.8in}
\begin{minipage}[b]{2.8in}
\caption{A cartoon of a very large nucleus in its rest frame.
In the labatory frame, Lorentz contraction causes all of the
quarks and nucleons to pile up at essentially 
the same value of $x^{-}$.
}
\label{Cartoon}
\end{minipage}
\hfill
\begin{minipage}[b]{2.8in}
\caption{Map of transverse momentum scales in the 
McLerran-Venugopalan model.
}
\label{Scales}
\end{minipage}
\end{figure}

Within the \MV\ model~\cite{papers1to4,paper9}, 
the Lorentz-contracted nucleus is treated 
as a pancake of color charge coming from the valence quarks.
For a large nucleus, this means that a tube of cross-sectional
area $d^2\xt$ will intercept a large number of valence quarks
(see Fig.~\ref{Cartoon}).  If we restrict ourselves to small
longitudinal momentum fractions, then all of the quarks in
the tube effectively overlap.  Because a large number of quarks
contribute, the color charge is in a high-dimensional representation
of the group, and may be approximated by a classical charge
density.  

To compute the gluon number density, we begin by solving the
classical Yang-Mills equation for the vector potential $A^{\mu}$
as a function of the color charge density $\rho$.  Next, the 
usual quantum mechanical average is approximated by
an average over an ensemble of nuclear charge distributions to compute 
two-point correlation
functions of the vector potential, $\cfun$.  These correlation
functions contain the gluon number operator; hence, they are be
connected to the gluon number density.  The result of 
the averaging process depends on the form chosen
for the two-point charge density correlation function
$\langle \rho^a(x^{-},\xt) \rho^b(x^{\prime -},\xprimet) \rangle$.
Finally,
we obtain the gluon number density by performing the
appropriate Fourier transform of the position space correlator.


Under what conditions is it legitimate to use the classical
approximation we have just described?  Schematically, the
situation is as illustrated in Fig.~\ref{Scales}.  If the typical
transverse momenta being considered is too small (below $\LQCD$),
then we are in the stong coupling regime, where we have no reason
to expect classical methods to be valid.  At larger transverse
momentum scales, we have weak coupling.  However, we cannot go
to too large a transverse momentum, or else the requirement that
we have a large number of quarks contributing in each patch $d^2\xt$
of nucleus no longer holds:  the quantum graininess begins to become
important.  At the upper end of where the classical approximation
is valid, the solution is QED-like, with a distribution function
which is proportional to $1/\qt\ts^2$.   Somewhere in between
is a cross-over scale where
the non-Abelian terms in the equations of motion become important,
and we require a solution which is summed to all orders.

We now turn to an examination of the basic form of the
classical solution, using the situation in electrodynamics as
our guide.
Consider a
point charge $e$ moving at the speed of light down the $z$
axis.  In this situation, the current is
\beq
J^{+} = e \ts\delta(x^{-}) \delta^2(\xt);
\quad J^{-} = 0; \quad \Jt = {0}.
\eeq
Unlike in the (more familiar) Lorentz gauge where the $\mu$th
component of the current generates the $\mu$th component
of the vector potential, in light cone gauge only the
transverse components of $A^{\mu}$ are non-zero:
\beq
A^{i}(x) = {{e}\over{2\pi}} \ts \Theta(x^{-}) \ts 
           {{x^i}\over{\xt\ts^2}}.
\eeq
This vector potential corresponds to the non-vanishing field tensor components
\beq
F^{+i} = - {{e}\over{2\pi}} \ts \delta(x^{-}) \ts 
           {{x^i}\over{\xt\ts^2}}.
\eeq
These components of the field tensor correspond to transverse
electric and magnetic fields of equal strength.
There are no longitudinal fields in this limit.
An observer sitting at some fixed position  $(\bt,x^3)$ would 
see no fields except at the instant when the charge
made its closest approach, when a $\delta$-function 
pulse would be seen.  The magnitude of the
pulse would be proportional to $1/b$.  The (unobservable) 
vector potential,
which was zero before the passage of the charge, takes on
a non-zero value for all times afterward.  Since the fields
vanish at these times, the late-time vector potential may be thought of
as some particular gauge transformation of the vacuum.

The overall features of the above description
continue to hold when we switch to QCD, although the
details are slightly altered by the presence of the non-Abelian
terms in the field equations.  The net effect of these terms is
to color-rotate the source in
a complicated fashion.  Nevertheless, the chromoelectric and
chromomagnetic fields are non-zero only at the instant of
closest approach by the charge, and the vector potential
switches from one gauge transform of the vacuum to a different
one at this instant.

We now outline the method used to compute the correlation
function $\cfun$.
Full details of this part
of the calculation are found in Ref.~\cite{paper9}.
Essentially, what one does is to expand the vector potential
in ``powers'' of the charge density $\rho$, and perform the
averaging by doing all possible pairwise contractions using
\beq
\langle \rho^a(x^{-},\xt) \rho^b(x^{\prime -},\xprimet) \rangle
= \delta^{ab} \ts\mu^2(x^{-}) \ts
\delta(x^{-}-x^{\prime -}) \ts\curlyD(\xt-\xprimet).
\label{RhoRho}
\eeq
In this expression $\mu^2(x^{-})$ is the color charge squared
per unit area per unit thickness ({\it i.e.}\ per unit $x^{-}$).  
We set up the calculation with a non-zero nuclear thickness
to avoid ambiguities in the commutator terms of the Yang-Mills
equations which would arise if we let $\rho \sim \delta(x^{-})$
exactly.  Refering back to Fig.~\ref{Cartoon}, we see that
the quarks in our ``tube'' of color charge typically come from 
different nucleons.  
Thus, we expect them to be uncorrelated,
hence the dependence $\delta(x^{-}-x^{\prime -})$ in Eq.~(\ref{RhoRho}).
The transverse dependence of the charge denisty correlator is
given by $\curlyD(\xt-\xprimet)$.  In Ref.~\cite{paper9},
it is argued that this should also be a delta-function,
since we are restricted to length scales $\lesssim \LQCD^{-1}$.
However, as we shall see, doing so is not consistent with color
neutral nucleons, and leads to severe infrared divergences in
the correlation funtion.

After doing all of the contractions, we resum the
series to obtain the master formula
\beq
\cfun =
{{\delta^{ab}}\over{N_c}}  \ts
{ 
{\partial_i \partial^{\prime}_j L(\xt-\xprimet)}
\over
{L(\xt-\xprimet)} 
} \ts
\biggl[
e^{ N_c \chi(x^{-},x^{\prime -}) L(\xt-\xprimet)}
- 1 \biggr].
\label{MVCorrelator}
\eeq
Eq.~(\ref{MVCorrelator}) depends on two new functions.
The first, $\chi(x^{-},x^{\prime -})$, 
measures the amount of charge in those layers of the source
which have already passed {\it both}\ of the points which we
are comparing:
\beq
\chi(x^{-},x^{\prime -}) \equiv 
\int_{-\infty}^{\min(x^{-},x^{\prime -})}
d\xi^{-}
\ts  \mu^2(\xi^{-}).
\label{ChiDef}
\eeq
The appearance of this function may be understood by
recalling that the value of the vector
potential depends on whether or not the charge has yet reached
its point of closest approach.  
Although the range of integration in~(\ref{ChiDef})
extends to $x^{-}=-\infty$,
in practice this is cut off by the form of $\mu^2(x^{-})$,
which for a pancake-shaped charge distribution, should be
non-zero only in a relatively small range near the value
of $x^{-}$ that corresponds to the position of the nucleus.

The second new function appearing in~(\ref{MVCorrelator})
is given by
\beqa
L(\xt-\xprimet) \equiv 
\int d^2 \xit \int d^2 \xiprimet \ts
\curlyD&&(\xit - \xiprimet) 
\Bigl[ G(\xt - \xit) G(\xprimet - \xiprimet)
\cr &&
-\half G(\xt-\xit) G(\xt-\xiprimet)
-\half G(\xprimet-\xit) G(\xprimet-\xiprimet) \Bigr],
\label{Ldef}
\eeqa
where
\beq
G(\xt-\xprimet) \equiv
{ {1}\over{4\pi} }
\ln \biggl( {
{\vert\xt-\xprimet\vert^2}
\over
{\lambda^2}
} \biggr).
\label{GreenFunction}
\eeq
This Green's function is the
inverse of the operator $\delt^2$ (in two dimensions).
The lack of a scale in our theory produces an
infrared divergence in this function, which is signalled
by the appearance of an arbitrary length scale $\lambda$.
Clearly, the infrared
finiteness or lack thereof of $L(\xt-\xprimet)$
is intimately related to the form chosen for 
$\curlyD(\xit-\xiprimet)$.
On the other hand, according to Eq.~(\ref{Ldef}),
$L(\xt-\xprimet)$ vanishes when $\xt=\xprimet$,
{\it i.e.}\ in the ultraviolet.
Thus, at very short distances 
the nonlinear terms in~(\ref{MVCorrelator}) drop out
and the behavior of the
correlation function is the same as if we had considered
a purely Abelian theory instead.

If, as in Ref.~\cite{paper9}, we take 
$\curlyD(\xit-\xiprimet) = \delta^2(\xit-\xiprimet)$,
we end up with
\beq
L(\xt-\xprimet) \sim {{1}\over{\delt^4}} 
\Bigl[ \delta^2(\xt-\xprimet) - \delta^2(0) \Bigr].
\eeq
Although the subraction term serves to remove the quadratic 
infrared singularity, a logarithmic divergence remains.
Assuming that the arbitrary scale should be of order $\LQCD$
on physical grounds, the authors of Ref.~\cite{paper9} obtain
\beq
\TRcfun =
{ {4(N_c^2-1)} \over {N_c\vert\xt-\xprimet\vert^2}} \ts
\biggl[ 1 - \Bigl( \vert\xt
            -\xprimet\vert^2 \LQCD^2\Bigr)^{(N_c/16\pi)
                \chi(x^{-},x^{\prime -}) 
                \vert\xt-\xprimet\vert^2}
\biggr].
\label{RunAway}
\eeq
This correlation function diverges like
$(\xt^2)^{\xt^2}$ for large values of the separation.
Of course, the bad behavior does not begin until the
point $x \sim \LQCD^{-1}$, the point where we begin to
mistrust our calculation anyhow.  Unfortunately, because of this
divergence, the Fourier transform of Eq.~(\ref{RunAway})
does not exist for any value of $\qt$, real or imaginary.
So it is difficult to see how to define the gluon number
density more than qualitatively using this expression.

The resolution of this problem lies in recognizing the
importance of enforcing the observation that nucleons, when
observed on a large enough length scale, should be color neutral.
A consequence of the Gaussian averaging employed in the \MV\ 
model is that the average color charge vanishes,
$\langle \rho^a(x^{-},\xt) \rangle = 0$.
However, we should also impose the (stronger) condition that
\beq
\int dx^{-} \ts d^2\xt \ts\ts \rho^a(x^{-},\xt) = 0
\label{Neutral}
\eeq
for a nucleus-sized volume.  If we integrate the charge density
correlator~(\ref{RhoRho}) over all $(x, x')$ and apply~(\ref{Neutral})
we obtain a constraint
on the transverse portion of the correlator $\curlyD$:
\beq
\int d^2\xt \ts\ts \curlyD(\xt) = 0.
\label{Key}
\eeq
Any correlation function which satisfies Eq.~(\ref{Key}) is
compatible with color neutral nucleons.  Such a correlation
function must contain an intrinsic scale, {\it i.e.}\ the
minimum transverse length scale for which~(\ref{Key}) becomes true.
When we compute $L(\xt)$, 
this scale will be imparted
to the logarithms appearing in Eq.~(\ref{Ldef}).
This stongly suggests that the
resulting correlation function determined from
Eq.~(\ref{MVCorrelator}) 
will be infrared finite.  Indeed, this is the case\cite{US}.

To illustrate the features of our improved treatment, we
turn to a specific model of a large nucleus introduced by
Kovchegov\cite{paper12}.  In this model, we view the nucleus
as containing $A$ nucleons of radius $a$ distributed within
a sphere of radius $R$.  Each ``nucleon'' consists of a $q\bar{q}$
pair.  By explicitly averaging over the allowed positions of the
quarks, antiquarks, and nucleons we may explicitly compute the
function $\curlyD(\xt-\xprimet)$.  We find that there are two types 
of terms.  The first is generated when the position of two
quarks (or two antiquarks) overlap.  It is proportional to
$\delta^2(\xt-\xprimet)$, precisely the form for $\curlyD$
employed in Ref.~\cite{paper9}.  The second kind of term
enters in with opposite sign and corresponds to the situation 
when a quark overlaps an antiquark.  It is proportional
to a smooth function of the separation.  Since  the focus
of Ref.~\cite{paper12} was on very short distances,
this term was neglected
relative to the delta-function contribution.
However, at somewhat longer distance
scales, it is precisely this additional contribution which 
is required to satisfy Eq.~(\ref{Key}).

\begin{figure}[t!] 
\includegraphics{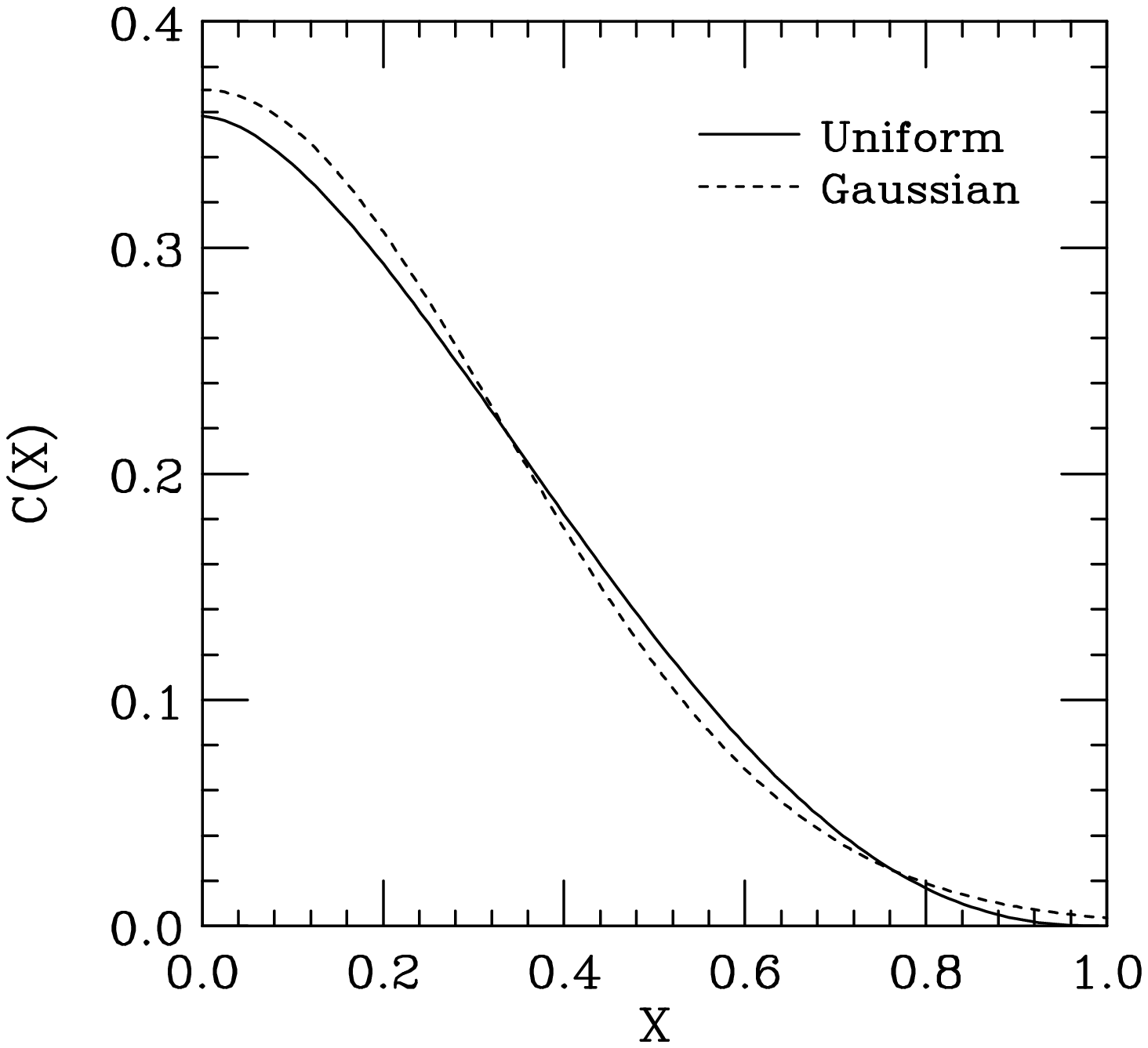}
\includegraphics{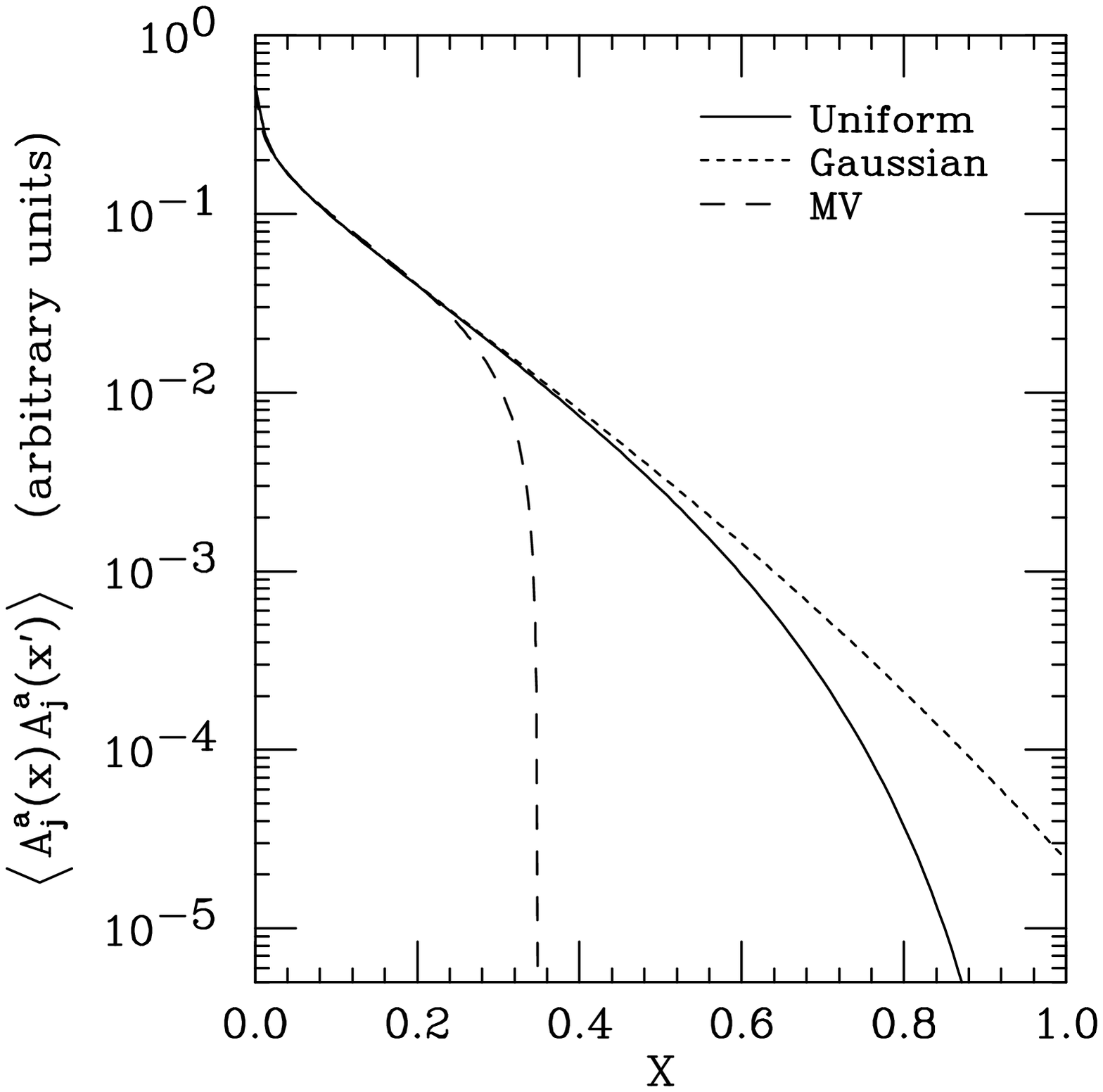}
\vspace{2.8in}
\begin{minipage}[b]{2.8in}
\caption{The smooth part of the two-point charge density 
correlation function in Kovchegov's model, defined by
writing $\curlyD(\xt) \equiv \delta^2(\xt) - C(\xt)$. 
The nucleon size parameters have been chosen so that the 
resulting gluon number densities match in the ultraviolet limit.  
The two curves correspond to
uniform and Gaussian distributions of the quarks inside
the nucleons.
}
\label{Ihat}
\end{minipage}
\hfill
\begin{minipage}[b]{2.8in}
\caption{The trace of the two-point vector potential correlation
function~(\protect\ref{MVCorrelator})  in position space.  
The nucleon size parameters $a$
and $\LQCD$  have been chosen so that the resulting gluon number
densities match 
for $X \rightarrow 0$.
We have fixed
the longitudinal coordinates at a place where $N_c \chi = 50 a^{-2}$.
Plotted are the results of Ref.~\protect\cite{paper9} (labelled ``MV'')
as compared to Kovchegov's model~\protect\cite{paper12}.
}
\label{TrCfun}
\end{minipage}
\end{figure}

\begin{figure}[t!] 
\includegraphics{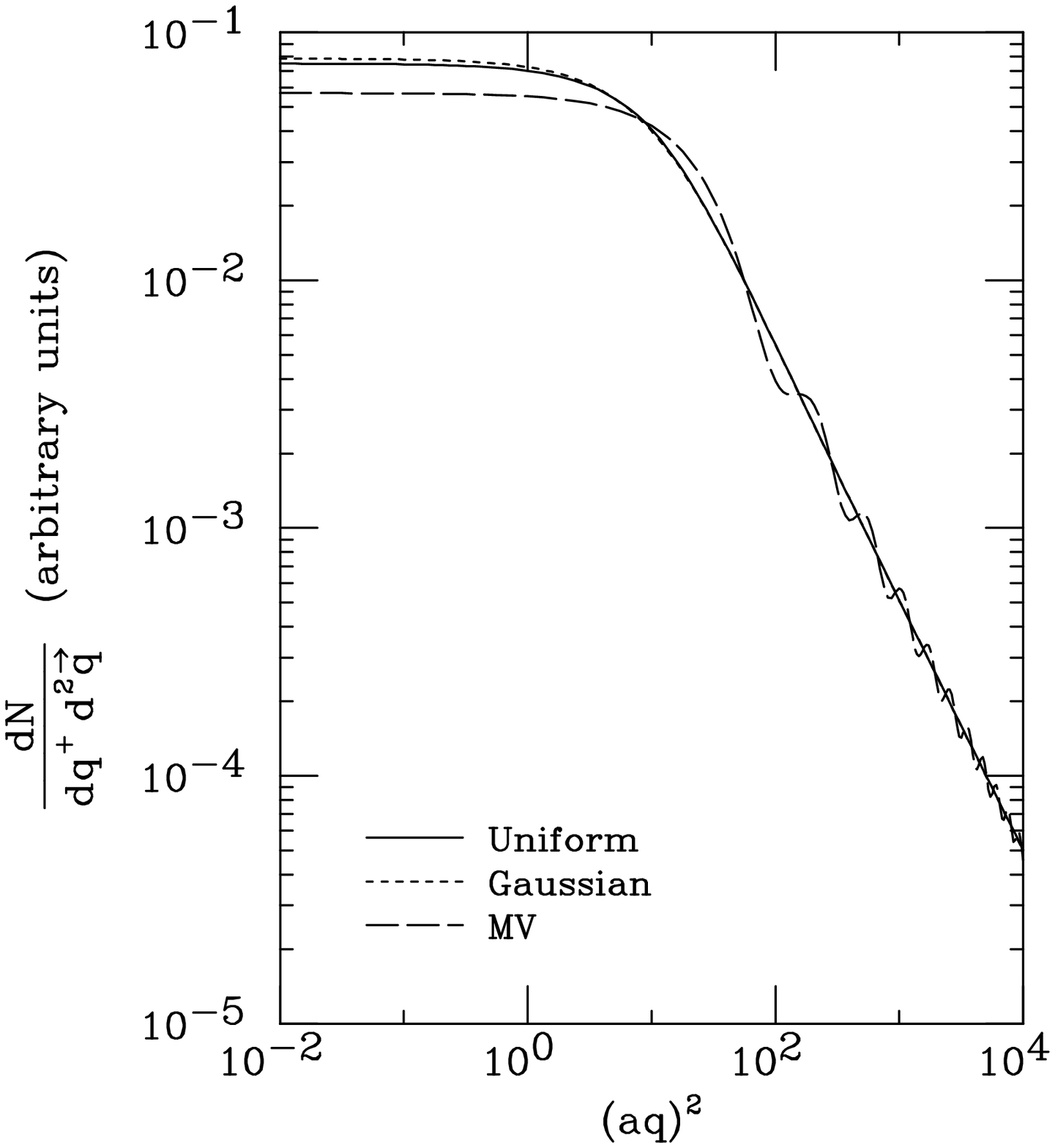}
\includegraphics{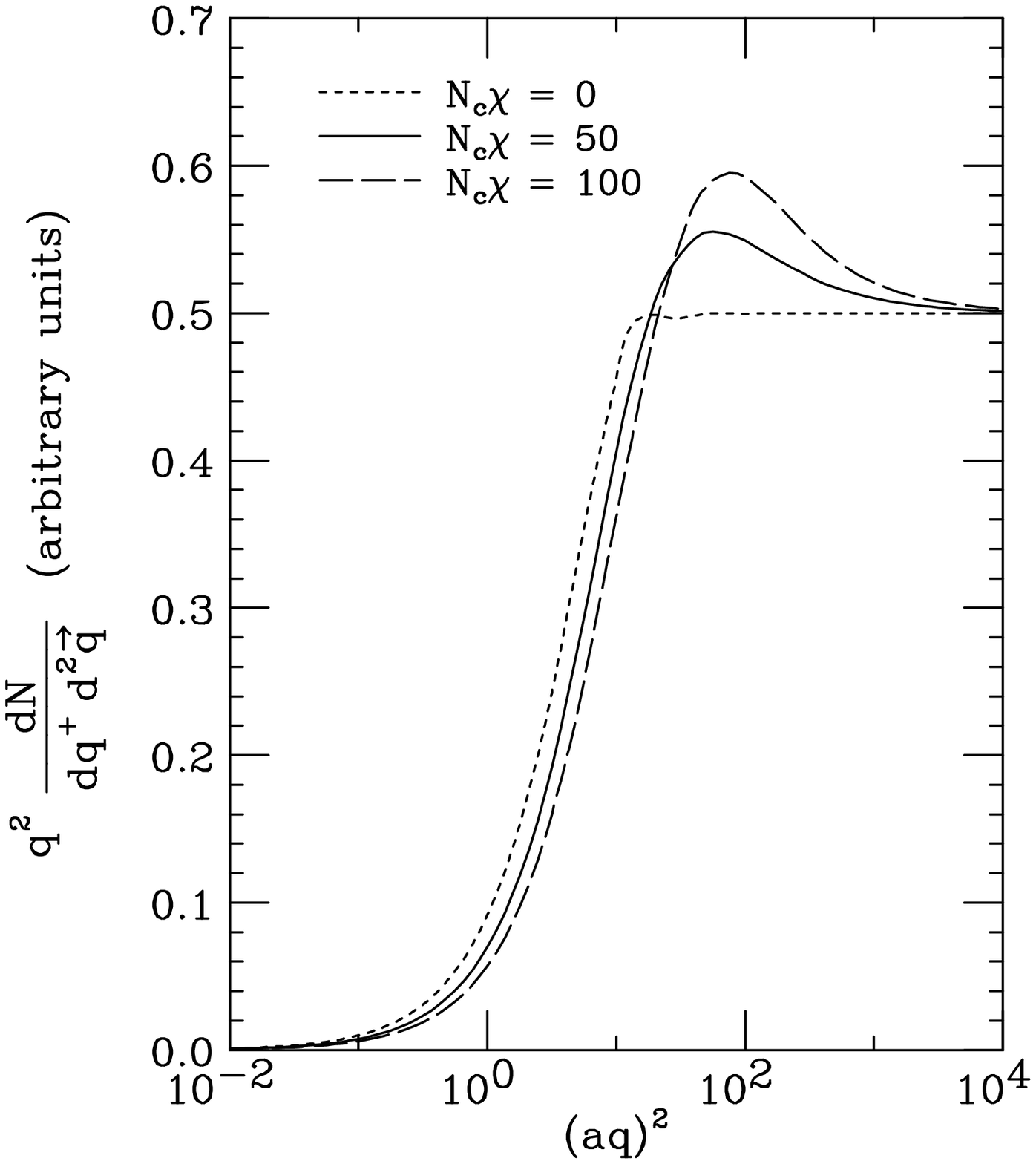}
\vspace{3.3in}
\begin{minipage}[b]{2.8in}
\caption{Plot of the gluon number density at fixed $q^{+}$.
The nucleon size parameters $a$ and $\LQCD$ have been
fixed so that these functions match for $\qt\ts^2 \rightarrow \infty$.
We have chosen $N_c \chi = 50 a^{-2}$.  
Plotted are
the results of Ref.~\protect\cite{paper9} (labelled ``MV'')
as well as results using Kovchegov's model~\protect\cite{paper12}.
}
\label{NumDen}
\end{minipage}
\hfill
\begin{minipage}[b]{2.8in}
\caption{Plot of $\qt\ts^2$ times
the gluon number density at fixed $q^{+}$ for
various values of the color charge density within
the uniform version of Kovchegov's model~\protect\cite{paper12}.
As $\chi$ increases, the importance of the non-Abelian 
contributions increases.  Conversely, $\chi \rightarrow0$
corresponds to the Abelian limit.
}
\label{HO}
\end{minipage}
\end{figure}

We have compiled a series of 
plots (Figs.~\ref{Ihat}--\ref{HO})
to aid in the comparison of our results using Kovchegov's model to
the results of Ref.~\cite{paper9}.
In addition to the uniform distribution
of quarks, antiquarks, and nucleons
employed by Kovchegov, we have also performed the averaging
using Gaussian distributions.
In preparing these plots, we have adjusted the nucleon
size parameter $a$ and (for the \MV\ result) $\LQCD$ so 
that the corresponding
correlation functions~(\ref{MVCorrelator}) match in the
ultraviolet limit.  This requires $a_G = 0.464 a_U$ and
$\LQCD = 1.44 a_U^{-1}$ where $a_U$ is the nucleon size
parameter for the uniform quark distribution.


In Fig.~\ref{Ihat} we have plotted the smooth part of the
correlation function for the uniform and Gaussian cases,
defined by writing 
$\curlyD(\xt-\xprimet) \equiv 
\delta^2(\xt-\xprimet) - \Ihat(\xt-\xprimet)$.
In both cases, the bulk 
of the contribution comes from
separations less than $2a$.\footnote{In fact, $\Ihat(\xt-\xprimet)$
vanishes identically for separations greater than $2a$
when a uniform distribution is assumed.}
Therefore, in this and subsequent plots 
we have defined the dimensionless
distance $X \equiv \vert\xt-\xprimet\vert /(2a)$.

Next, we present
Fig.~\ref{TrCfun}, which shows the trace of the correlation 
function~(\ref{MVCorrelator}) for all three cases.
The Fourier transform of this quantity is proportional to
the gluon number density.
The Gaussian curve, not surprisingly, has a Gaussian tail
for $X \ge 2a$, while the uniform curve vanishes identically
in this region.  On the other hand,
the result from Ref.~\cite{paper9} runs off to $-\infty$ 
beyond $\LQCD^{-1}$.

We come to the gluon number density in Fig.~\ref{NumDen}.
To define the \MV\ curve, we have followed the suggestion 
made in Ref.~\cite{paper10}, and simply cut off the $\xt$
integration at $x = \LQCD^{-1}$.  This is the source of the
wiggles visible in Fig.~\ref{NumDen}.  All three curves
have the same overall shape, with a plateau at small values
of $\qt$ and a $1/\qt\ts^2$ fall off at large $\qt$.  The number
of zero momentum gluons clearly depends on how we have cut
things off:  the Gaussian model, which allows quarks to be
(albeit with small probability) a large distance from the center
of their nucleons has the most long wavelength gluons, where
as the \MV\ curve, generated with a hard cutoff, has the fewest.

Finally, we display the effect of the non-Abelian terms on
the gluon number density.  This is most easily seen on a 
plot of $\qt\ts^2$ times the number 
density (Fig.~\ref{HO}).\footnote{The fact that this
curve approaches a constant for large $\qt\ts^2$ demonstrates
the $1/\qt\ts^2$ behavior described in the text.}
From Eq.~(\ref{MVCorrelator}), it is clear that the magnitude
of $N_c \chi$ governs the importance of these contributions.
In Fig.~\ref{HO} we have taken $N_c \chi$ to 
be 0, $50a^{-2}$, and $100a^{-2}$.
What we see in this plot is a transfer of gluons from low
values of $\qt$ to higher ones.  In fact, it is possible
to show that the area under this plot is conserved\cite{US},
and that this shifing of gluons from one energy to another
is the only effect of the non-Abelian contributions.

In conclusion, we see that imposing a color-neutrality condition
on the nucleons eliminates the divergent infrared behavior 
of the two-point vector potential correlation function
in the \MV\ model.  
Because we have obtained a well-defined
expression for the gluon number density, we are able
to perform a quantitative investigation of the features
of the \MV\ model.  For example, we have shown that the \MV\ model
predicts a  gluon number density which is proportional to $1/q^{+}$
to all orders in the charge density.  We are
able to compute the gluon structure function in this framework,
including its absolute normalization.  The details of these
and other related results may be found in Ref.~\cite{US}.


\end{document}